# Minimum thermal conductance of twisted-layer graphite nanofibers

**Van-Truong Tran[1*], Thanh-Tra Vu[2], Philippe Dollfus[1], Jérôme Saint-Martin[1], and Marco Pala[1]**

[1]Université Paris-Saclay, CNRS, Centre de Nanosciences et de Nanotechnologies, 91120, Palaiseau, France.
[2]Department of Physics, School of Education, Can Tho University, Can Tho, Vietnam

Email: [*]vantruongtran.nanophys@gmail.com

**Abstract**

We study the thermal transport properties of twisted-layer graphite nanofibers. We show that in the presence of a twisted layer, the phonon thermal conductance of a graphite nanofiber varies remarkably with the twisted angle and can reach minimum values either at two critical angles $\theta_1$ and $\theta_2$ that conform to the rule $\theta_1 + \theta_2 = 180^0$ or exactly at the angle $\theta = 90^0$. A reduction of roughly 50% of the phonon thermal conductance can be achieved in some structures. We unveil that the twisting effect mainly influences the optical modes, leaving almost unaltered the acoustic ones. The effect is also visible in the higher and more numerous van Hove singularities of the phonon density of states. We also point out that the behavior of the thermal conductance with the twisted angle is associated with and dominated by the alteration in the overlap area between the twisted and non-twisted layers. The finite-size effect is demonstrated to play an essential role in defining the critical angles at the local minimums, where these angles are dependent on the size of the investigated nanofibers, in particular on the proportion between the widths of zigzag and armchair edges.

## 1. Introduction

The twisting effect in bilayer graphene results in novel physics when compared with non-twisted structures[1–5]. In particular, exceptional electronic characteristics with superconducting states have recently been reported at the magic angle of around 1.1 degree[6]. Numerous researches on the electronic properties of twisted structures in bilayer and few-layer graphene systems[7–12] as well as other 2D materials like black phosphorus[13] have been motivated by this fascinating physics. It also creates a new field of electronics known as "twistronics" and opens up a vast of opportunities of uses of the twisting effect[14].

The findings on the electronic properties of the twisted graphene layers have also inspired studies on the impact of this effect on the thermal properties of structures. However, the number of studies on the thermal properties of twisted graphene structures is still modest[15,16].

Regarding in-plane phonon characteristics, in 2013 Cocemasov *et al*.[17] studied phonon properties in twisted bilayer graphene, focusing more on AA-stacking structures. They showed that the twisting effect induces the largest impact on out-of-plane acoustic (ZA) modes in the phonon bands of the bilayer structures. Even so, the impact is still noticeable in optical and other acoustic modes[17], and also in AB-stacking bilayer structures[18]. Interestingly, both experiment and molecular dynamic (MD) simulations showed in-plane phonon conductivity reaching the minimum near the twisted angle equal to $15^0$ or $45^0$ and a local maximum at $30^0$[16,19], demonstrating tunable features of thermal transport in twisted bilayer and multiple layer graphene systems.

Although for few-layer graphene structures, in-plane phonon transport is more crucial, thermal transport across the layers was also assessed. By using MD simulations, Wang *et al*.[15], and Nie *et al*.[16] pointed out that the perpendicular transport thermal conductivity in bilayer and few-layer graphene structures could reach a minimum at the twisted angle of about $30^0$, in contrast to that of the in-plane thermal transport[16,19].Such results are essentially observed in 2D structures, i.e., the structures with layers are assumed to spread infinitely, while the investigation of the thermal properties of twisted structures with finite sizes is still lacking.

It is worth noting that the finite-size effect in structures such as quasi-1D graphene ribbons offers unique physics compared to those in 2D structures. For instance, the observed finite-bandgap in these structures as well as novel physics from the edge-type terminations[20] make them suitable for a variety of electronic applications[20–22]. Thus one could expect that the combination of both twisting and finite-size effects might result in novel phenomena in finite-size multiple-layer graphene structures compared to what has been observed in 2D counterparts. Such studies still need to be conducted at both fundamental and practical levels.

Graphite nanofibers (GNFs) have recently attracted great attention for their extremely high thermoelectric capacity[23]. Such structures can be also considered as relevant systems for exploring the mutual twisting and finite-size effects due to the high probability of twisted layers present in the systems and the fact that each basic cell has finite sizes terminated by zigzag and armchair edges. Therefore, in this work, we focus on the impact of the twisting effect on the thermal transport in twisted-layer graphite nanofibers. Our study aims not only at providing a fundamental understanding of the twisting effect in finite-size structures but also intends to reveal how such an effect could play a role in applications of graphite nanofibers.

The paper is organized as follows. In Sec. 2, we describe in detail the studied systems and then briefly present the methodology based on the efficient Force Constant (FC) model and the Non-Equilibrium Green's Functions (NEGF) formalism. Sec. 3 discusses the obtained results, in which in Sec. 3.1, we investigate the variation of phonon thermal conductance as a function of the twisted angle, and in Sec. 3.2, we focus on analyses of the role of finite-size effect. Sec. 4 is devoted to conclusions.

## 2. Studied system and methodology

### 2.1. Studied system

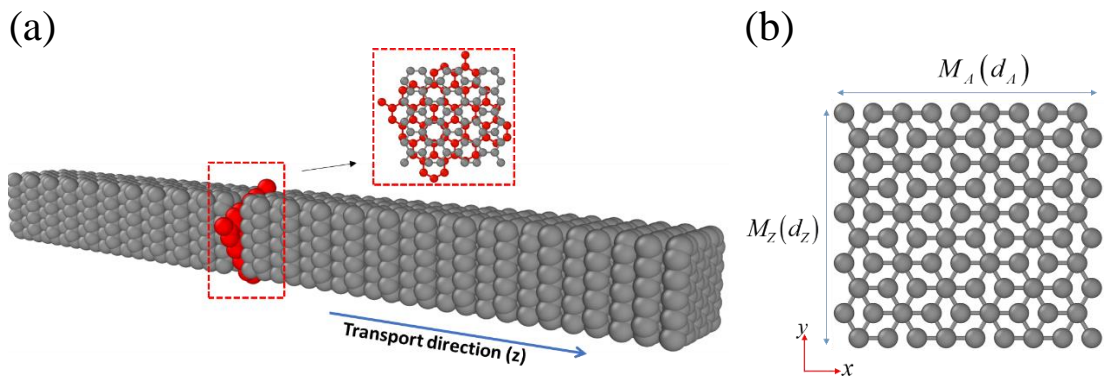

**Figure 1**. (a) Sketch of a nanodevice made of "platelet" graphite nanofibers with a twisted layer marked in red. (b) The atomistic view of a basic cell containing $2M_A \times M_Z$ atoms. The illustration is for the nanofibers with $M_A = 6$ ($d_A \approx 1.207$ nm), $M_Z = 10$ ($d_Z \approx 1.107$ nm).

Fig. 1(a) illustrates the studied device made of a GNF in the presence of a twisted layer that is marked in red. The transport direction is along the c-axis, perpendicular to the graphite layers. It is worth mentioning that among several types of GNFs, this kind of structure is classified as "platelet" graphite nanofibers[24,25]. For the sake of convenience, the c-axis is identical to the z-axis and the rotation is performed around this axis. In Fig. 1(b), a basic cell containing two sub-layers is shown. The diameter of the fiber is characterized by the number of slices in each sub-layer along the armchair and zigzag edges, $M_A$ and $M_Z$, respectively. The widths $d_A$ and $d_Z$ of the GNF along armchair and zigzag edges can be determined as $d_A = (3M_A - 2) \times a_0 / 2 + a_0 / 2$ and $d_Z = (M_Z - 1) \times \sqrt{3}/2 \times a_0$, where $a_0 \approx 0.142$ nm is the distance between the two nearest in-plane carbon atoms.

In this work, we consider the nanofiber with diameters from sub-nano to more than 2 nm, which are feasible with the recent development of the CVD technique in growing carbon structures[26].

## 2.2. Methodology

The basic cell of studied graphite nanofibers contains hundreds of atoms, in addition, we aim to investigate the phonon transport properties in defect twisted structures, these systems are highly challenging with *ab initio* calculations. We, therefore, employed Force Constant (FC) model as an efficient approach to construct dynamical matrices, and combined with the Non-Equilibrium Green's Functions (NEGF) formalism for simulating the transport properties of phonons.

### *Force Constant model*

The FC model containing both in-plane and van der Waals (vdW) interaction was employed, which accurately reproduces the phonon dispersion of both graphene layers and graphite along the c-axis obtained by *ab initio* methods and experimental measurements [18,27]. The secular equation for phonons deriving from Newton's second law is written [28,29]:

$$\omega^2 U = DU, \quad (1)$$

where $U$ is the column matrix containing the amplitude vectors of vibration of all the lattice sites, $\omega$ is the angular frequency, and $D$ is the Dynamical matrix which is calculated as [28]:

$$D = \left[ D_{3\times 3}^{ij} \right] = \left[ \begin{cases} -\dfrac{K_{ij}}{\sqrt{M_i M_j}} & \text{for } j \neq i \\ \sum_{j \neq i} \dfrac{K_{ij}}{M_i} & \text{for } j = i \end{cases} \right], \quad (2)$$

with $M_i$ is the mass of the *i*-th atom and takes the value of $1.994 \times 10^{-26}$ kg for each carbon atom. $K_{ij}$ is the 3×3 tensor coupling between the *i*-th and *j*-th atoms and is defined depending on whether it is in-plane or inter-plane interaction, i.e.,

(i) For in-plane interactions, $K_{ij}$ was defined by a unitary in-plane rotation[28,29]

$$K_{ij} = U^{-1}(\theta_{ij}) K^0_{ij} U(\theta_{ij}), \qquad (3)$$

In which $U(\theta_{ij})$ is the rotation matrix [29] that is defined by

$$U(\theta_{ij}) = \begin{bmatrix} \cos(\theta_{ij}) & \sin(\theta_{ij}) & 0 \\ -\sin(\theta_{ij}) & \cos(\theta_{ij}) & 0 \\ 0 & 0 & 1 \end{bmatrix} \qquad (4)$$

$\theta_{ij}$ is the anticlockwise rotating angle formed between the positive direction of the *x*-axis and the vector from the *i*-th atom to the *j*-th atom. $K^0_{ij}$ is the force constant tensor given by:

$$K^0_{ij} = \begin{pmatrix} \Phi_r & 0 & 0 \\ 0 & \Phi_{t_i} & 0 \\ 0 & 0 & \Phi_{t_o} \end{pmatrix}, \qquad (5)$$

where $\Phi_r, \Phi_{t_i}$ and $\Phi_{t_o}$ are the force constant coupling parameters in the radial, transverse in-plane, and out-plane directions, respectively. In this work, these in-plane force constant parameters were considered up to four nearest neighbor interactions, and therefore twelve parameters are needed for in-plane couplings. The values of these parameters were taken from Wirtz's work [27].

(ii) For interactions between layers (vdW interactions), we employed the spherically symmetric interatomic potential model, in which each component of the coupling tensor $K_{ij}$ is defined by:[18]

$$(K_{ij})_{kk'} = \delta(r^{ij}) \frac{r^{ij}_k r^{ij}_{k'}}{(r^{ij})^2} \qquad (6)$$

in which $k, k'$ is one of the *x*, *y*, *z* components. $r^{ij}$ is the vector from the *i*-th to the *j*-th atoms. $\delta(r^{ij})$ is the decaying component $\delta(r^{ij}) = A.\exp(-r^{ij}/B)$ with empirical parameters $A = 573.76$ N/m, $B = 0.05$ nm. It is worth mentioning that the minus sign "-" is not present in equation (6) as in ref. [18] because this sign has been included in equation (2). In practice, to have the best fit between the FC model and the experimental data for bulk graphite, we chose the distance between two graphite layers equal to 0.328 nm. In addition, to simplify the computation, a cut-off of 1 nm in real space was applied for $r^{ij}$ in Eq. (6).

### *Green's function formalism for transport study*

To investigate the phonon transport properties, we employed the NEGF technique which is highly relevant to study transport in nanostructures, including defects [30]. Within this technique, all considered structures are divided into three parts: the left and right leads and the device (central) region. The leads were considered as semi-infinite periodic regions. The device region contains studied defects and has the length characterized by the number of basic cells $N_A$.

The retarded Green's function for phonon can be written as:

$$G = \left[ \omega^2 + i\eta - D_D - \Sigma^s_L - \Sigma^s_R \right]^{-1}, \tag{7}$$

where $\eta$ is a positive infinitesimal number, $D_D$ is the dynamical matrix of the device (scattering region), and

$$\begin{aligned} \Sigma^s_L &= D_{DL} G^0_L D_{LD} \\ \Sigma^s_R &= D_{DR} G^0_R D_{RD} \end{aligned} \tag{8}$$

are the surface self-energies contributed from the left and right contacts. $G^0_{L(R)}$ is the surface Green's function of the isolated left (right) contact. The self-energies were computed based on the Sancho-Rubio iterative technique[31].

To compute efficiently phonon transmission, the recursive technique [32] was employed to handle the size of the device Hamiltonian in the Green's function calculation. Phonon transmission was then computed as [28,30]:

$$T_p = Trace\left\{ \Gamma^s_L \left[ i\left(G_{11} - G_{11}^\dagger\right) - G_{11} \Gamma^s_L G_{11}^\dagger \right] \right\}, \tag{9}$$

where $\Gamma^s_{L(R)} = i\left(\Sigma^s_{L(R)} - \Sigma^{s\,\dagger}_{L(R)}\right)$ is the surface injection rate at the left (right) contact.

The phonon thermal conductance $K_p$ was computed by [28]

$$K_p = \frac{k_b}{2\pi} \int_0^\infty d\omega \mathrm{T}_p(\omega) g^p(\omega, T), \tag{10}$$

where $g^p(\omega,T) \equiv \left(\dfrac{\hbar\omega}{2k_B T}\right)^2 / \sinh^2\left(\dfrac{\hbar\omega}{2k_B T}\right)$.

## 3. Results and discussions

In this section, we first examine the variation of the phonon thermal conductance as a function of the twisted angle. Then, we analyze the role of the finite-size effect in defining the characteristics of such variation.

### 3.1. Crucial angles at minimum thermal conductance

We investigate the dependence of the thermal conductance of the studied nanofiber on the twisted angle of a single layer within the system.

Fig. 2(a) shows the phonon thermal conductance at room temperature as a function of the twisted angle of the GNF with the size $M_A = 6$ ($d_A \approx 1.207$ nm), $M_Z = 10$ ($d_Z \approx 1.107$ nm). The results of three cases with different positions of the twisted layer and different device lengths are displayed. The position of the twisted layer is charactered by two indices [$n_{cell}$, $n_{layer}$] which indicates the $n_{cell}$-th cell in the device region that contains the twisted layer, and the twisted layer is the layer 1 or 2 among the two layers in a basic cell.

For the twisted angle varies from 0 to $180^0$, at the first glance, it can be seen that the phonon thermal conductance varies remarkably with the twisted angles. Interestingly, the variation forms two valleys around $45^0$ and $135^0$ together with a peak located around the twisted angle of $90^0$. It is worth mentioning that with twisted angles from $180^0$ to $360^0$, the results (not shown) lead to a symmetrical curve with the one from 0 to $180^0$. Such results make sense since rotations from $180^0$ to $360^0$ are equivalent to those from 0 to $-180^0$, which is symmetrical with rotations from 0 to $180^0$. On the other hand, there is not a perfect symmetry around $90^0$ as it induces different stackings. For example, the non-twisted structure and the structure twisted with an angle of $180^0$ do not have the same conductance as they exhibit AB and AA stackings.

It can be observed although the minimum thermal conductance depends on the position and the length of the device, the twisted angle corresponding to the minimum thermal conductance remains unchanged, i.e., at $45^0$ and $135^0$ for the studied structure. In addition, these two angles incidentally satisfy the rule

$$\theta_1 + \theta_2 = 180^0 . \tag{11}$$

The thermal conductance of twisted structures is reduced remarkably compared to that of the non-twisted one. For example, in the case of the green curve, at $45^0$ and $135^0$, $K_p$ reaches a minimum at the values of 0.148 nW/K and 0.147 nW/K, respectively and compared to the conductance of 0.229 nW/K of the non-twisted structure, it thus reduces of about 35% in the value of the phonon thermal conductance. Such a reduction in the thermal conductance is very appealing for thermoelectric applications since it can boost the thermoelectric performance of the graphite nanofiber.

The phonon thermal conductance as a function of the twisted angle is given in Fig. 2(b) at temperatures other than room temperature. It shows clearly that the variation at different temperatures is similar. Interestingly, the twisted angles at the minimum thermal conductance are likewise unaltered.

The results depicted in Figs. 2(a) and 2(b) thus show that the two twisted angles at the minimum conductance are crucial and independent of the position of the twisted layer as well as of temperatures.

(a)

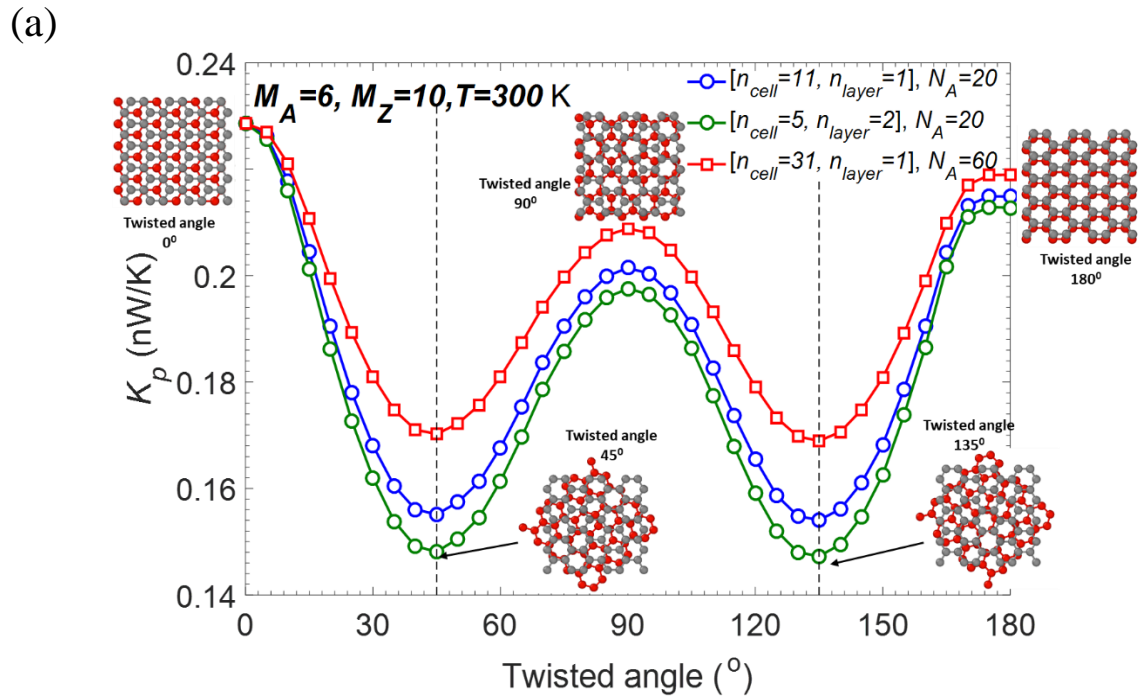

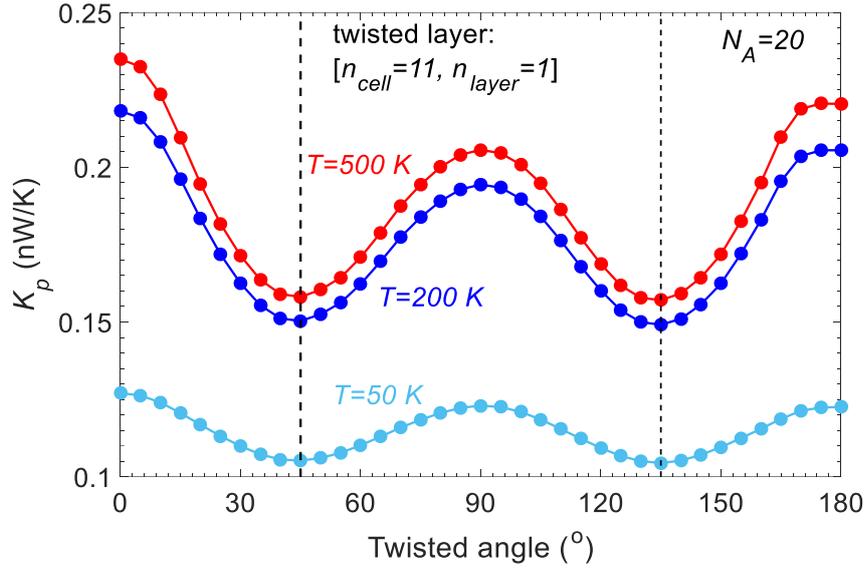

**Figure 2**. (a) Thermal conductance at room temperature is plotted as a function of twisted angle with different positions of the twisted layer and different device length $N_A$. (b) Thermal conductance as a function of the twisted angle at different temperatures. In all cases, the same size nanofiber was considered $M_A = 6$, $M_Z = 10$.

To understand the physics underlying the variation in the phonon thermal conductance in the presence of a twisted angle, it is relevant to look at the phonon bands of the periodic structures even though defects do not appear periodically in the systems. For twisted structures, we will consider periodic structures with a twisted layer in a basic cell. The phonon bands can be defined from the eigenvalues of the equation which is simplified from Eq. (1) to the equation for a single unit cell:

$$d\mathbb{U} = \omega^2 \mathbb{U}, \tag{12}$$

where $d$ is the dynamic matrix defined as

$$d = D_{00} + \sum_{\alpha \neq 0} D_{0\alpha} \cdot e^{i\vec{k}\cdot(\vec{R_\alpha}-\vec{R_0})}, \tag{13}$$

With $D_{00}$ is the dynamical matrix of the cell 0 and $D_{0\alpha}$ is the dynamical matrix presenting the coupling between the cells 0 and $\alpha$.

The influence of the twisting effect on the phonon modes can be confirmed using the phonon density of states (DOS), which is very sensitive to changes in the phonon bands. We calculated the phonon DOS by using the Gaussian smearing of the Dirac delta function [33], i.e.,

$$DOS(\omega) = \sum_n \sum_{\vec{k} \in BZ} \delta\left(\omega - \omega_n(\vec{k})\right) \approx \sum_n \sum_{\vec{k} \in BZ} \frac{1}{\eta\sqrt{\pi}} e^{-\frac{(\omega - \omega_n(\vec{k}))^2}{\eta^2}}, \quad (14)$$

in which $n$ is the phonon band index, $\vec{k}$ is a wave vector in the first Brillouin zone (BZ), $\eta$ is a small positive number, and $\omega_n(\vec{k})$ is the frequency of the $n$-th phonon mode at the wave vector $\vec{k}$, which is calculated from equation (12). Eq. (14) is a generic formula that can also be used to calculate the phonon DOS for 2D and 3D structures.

Phonon bands for three cases of the twisted angle are shown in Fig. 3(a). As can be seen, when the structure is twisted, mini-gaps are slightly widened between optical bands and, in parallel, these bands are flattened. These effects seem stronger with a greater twisted angle. However, it appears that the acoustic modes are only marginally affected by the distortion of the structure. Such an outcome reveals that the effect of the twisted layer is similar to that of isotope doping observed in in-plane graphene ribbons[34].

The DOS depicted in Fig. 3(b) makes this insight more obvious. The peaks in DOS present van Hove singularities that are particularly noticeable around nearly flat bands. The DOS of twisted structures (color lines) shows more numerous as well as higher peaks compared to that of the non-twisted structure (black). Also, the twisted structures present larger mini-gaps as well as open new mini-gaps right above 100 cm$^{-1}$. The effect is observed more strongly in the frequency range from 70 to 250 cm$^{-1}$. Below 50 cm$^{-1}$ where acoustic modes locate, the DOS is weakly modified. These results are consistent with what has been observed with the phonon bands. The DOS of the case twisted angle equal to $60^0$ was also added for comparison. It manifests that the DOS of the structure with twisted angle $60^0$ presents smaller changes than that of the case of $45^0$. This reflects clearly around the mini-gaps and in the frequency range from 70 to 150 cm$^{-1}$. This outcome is in line with the transport characteristics seen in Fig. 2 for the phonon thermal conductance and suggests that the twisting seems to have the highest impact around $45^0$.

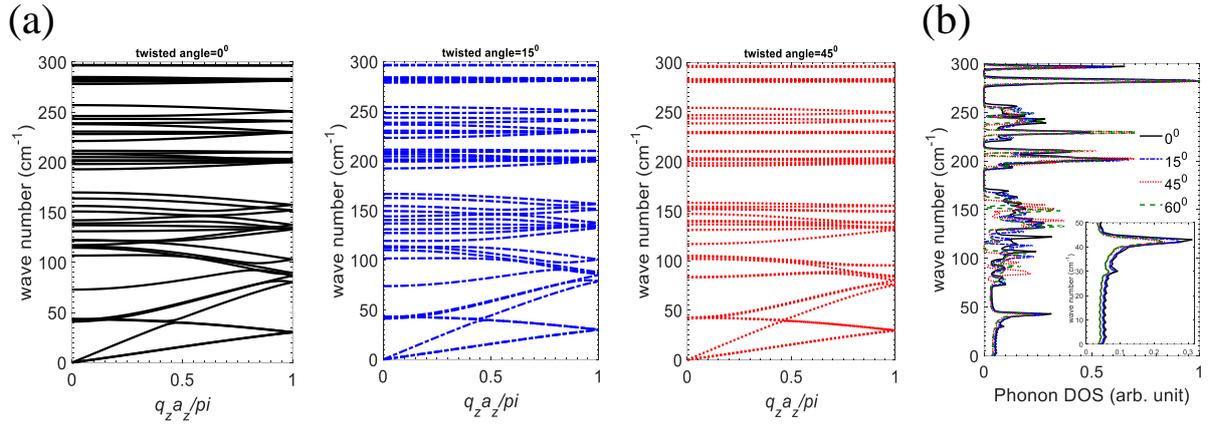

**Figure 3**. (a) Phonon bands of non-twisted and twisted periodic structures with an assumption that a layer in the basic cell is rotated. (b) Phonon density of state corresponding to different twisted angles. The size of the nanofiber: $M_A = 6$, $M_Z = 10$.

Although the phonon bands and the phonon DOS give some insights to interpret the transport properties, it is challenging to exploit them to monitor change at various twisted angles.

To understand the factor linked to the variation of the thermal conductance at each twisted angle, we focus on the overlap area between the twisted and non-twisted layers, as shown in atomic structures inserted in Fig. 2(a) for different angles. At the twisted angle of $45^0$, the overlap area is smaller than that at $0^0$ and $90^0$, which implies that the variation of the phonon thermal conductance might be associated with the change of the overlap area between the twisted and non-twisted layers. To elucidate this prediction, we explored the overlap area by considering two layers as two polygons and examined the overlap of these polygons when one is twisted. The overlap area as a function of the twisted angle is shown in Fig. 4.

Interestingly, the variation of the overlap area with the twisted angle in Fig. 4 is very similar to that of the phonon thermal conductance in Fig. 2. The two local minima are also located at $45^0$ and $135^0$. Thus, the main characteristics of the curve of $K_p$ against the twisted angle are strongly associated with the overlap area of the twisted and non-twisted layers. It is worth noting that, the overlap area at $90^0$ is seen to be close the that at $0^0$, however, the thermal conductance at these two angles is visibly distinct as can be seen in Fig. 2(a). This indicates that the thermal conductance of the twisted layer depends also on the relative position between atoms in different layers, which is understandable as it defines the strength of the van der Waals interactions. However, the remarkable resemblance between the variation of the overlap area and the thermal conductance with the twisted angle reveals that the overlap area between twisted and non-twisted layers in nanofibers dominates the characteristics of the change in the thermal conductance.

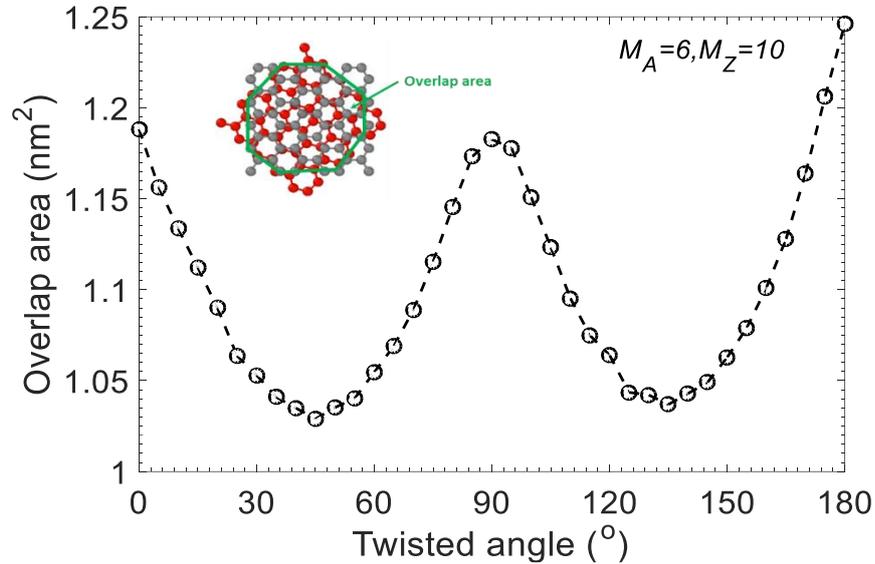

**Figure 4**. The overlap area of the two layers in a basic cell when one layer is twisted. Here $M_A$ = 6, $M_Z$ = 10.

### 3.2. Finite-size effect

In general, the size of the nanofibers, namely $d_A$ and $d_Z$, determines the overlap area between the twisted and non-twisted layer. On the other hand, as demonstrated in Sec. 3.1, a correlation between the overlap area and the thermal conductance exists, thus suggesting that the crucial angles at the minimum thermal conductance might be subjected to the finite-size effect. In this section, we, therefore, consider the variation of thermal conductance with the twisted angle for nanofibers of different sizes to unveil the role of this effect.

In Fig. 5, the thermal conductance at room temperature of four different size twisted structures is shown. Due to the significant difference in the thermal conductance across structures of various sizes, to make an efficient observation of global characteristics of all curves, we normalized the obtained thermal conductance of the twisted structure to that of the non-twisted one for each structure size considered.

As can be observed, most structures exhibit a curve with two valleys at the two sides of a peak located at the twisted angle of $90^0$, similarly to those observed for the structure $M_A = 6$, $M_Z = 10$ shown in Fig. 2(a). However, the depth and the position of the valleys depend on the size of the structure. Interestingly, the curve corresponding to the structure with the size $M_A = 6$, $M_Z = 7$ (blue diamonds) displays a distinct pattern in relation to the twisted angle, i.e., the lowest thermal conductance is precisely determined at $90^0$.

It is also worth noting that the largest considered structure (cyan triangles) shows a reduction of thermal conductance could reach almost 50%, which is greatly interesting for applications such as thermoelectrics.

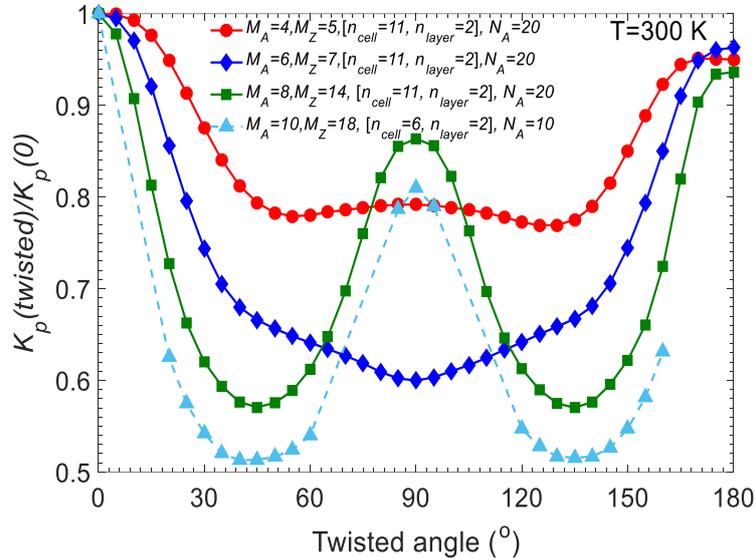

**Figure 5**. Phonon conductance of twisted nanofibers with different sizes normalized to that of the non-twisted structures.

To explore the finite-size effect, we first examine the crucial angle at the first local minimum $K_p$ as a function of the crossing area $S = d_A \times d_Z$. The results are shown in Fig. 6(a). It is important to note that, in all instances, Eq. (11) remains true for the two twisted angles at the lowest thermal conductance, therefore an investigation at the first important angle should be sufficient.

Fig. 6(a) clearly shows that the first essential angle varies significantly with the crossing area of the nanofibers. The black lines at each data point present the error bar. Due to the computational burden, in our simulations, we took the step of the twisted angle is $5^0$, thus the error bar for each side (upper and lower) of a data point is $2.5^0$. The crucial angle of the bigger structure appears to be smaller at the first look. However, the structure $M_A = 6$, $M_Z = 7$ (blue symbol) has an abnormal larger critical angle than that of the structure $M_A = 4$, $M_Z = 5$ (red symbol), despite having a larger crossing area. The crucial angle depends also on other factors than only the crossing area.

We notice that the three largest structures considered here have $d_A/d_Z \sim 1$ displaying a small difference in the crucial twisted angle (around $45^0$ and $40^0$), while the two structures which have a higher crucial angle, possess the size with $d_A/d_Z$ is far from 1. To unveil the role of the ratio $d_A/d_Z$ (we can consider also $d_Z/d_A$), we plot the crucial angle of the first minimum conductance as a function of the ratio $d_A/d_Z$ in Fig. 6(b). As can be observed, the dependence of the crucial angle on the proportion of the edge widths presents a clearer rule than that with the crossing area. With a ratio around 1, the crucial angle is seen around $40^0$-$45^0$ and can be converged to lower values if the crossing area is larger. When this ratio is far from 1, it unveils that the crucial angle increases further, and can reach the value of $90^0$, if the ratio is significantly large, above 1.6.

Thus, the crucial angle is remarkably size dependent varies more strongly with the difference between the two typical edge widths.

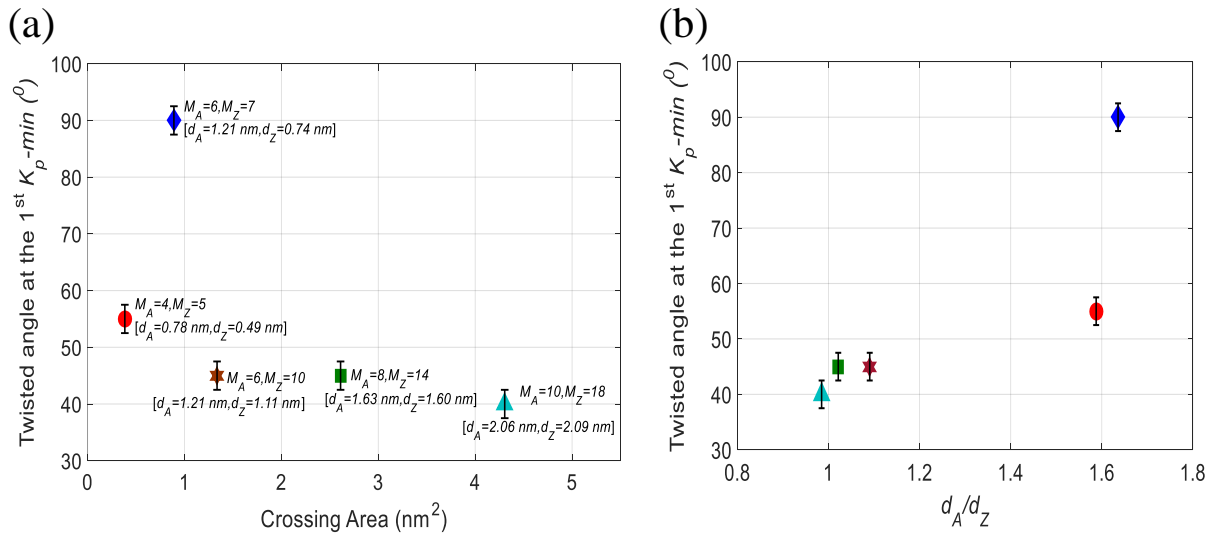

**Figure 6**. The crucial twisted angle at the first local minimum thermal conductance of different size nanofibers is plotted as a function of (a) the crossing area $S = d_A \times d_Z$, and (b) of ratio $d_A/d_Z$.

The three red, green, and cyan curves with symbols in Fig. 5 display the variation is similar to that of the structure shown in Fig. 2. We checked and observed the same behavior for the overlap area for these structures, i.e., being comparable to what is depicted in Fig. 4.

The blue curve with symbols in Fig. 5 shows a distinct behavior, and it is worth checking if the variation of the overlap area still agrees with that of the thermal conductance.

In Fig. 7(a), we show the thermal conductance at room temperature for two cases of different positions of the twisted layer. It can be seen clearly that the minimum conductance in the two curves occurs at the same twisted angle of $90^0$. The atomistic view around the twisted layer for some values of the twisted angle is also inserted in the panel and shows that the smaller overlap area seems to be observed around $90^0$. To confirm this, we examine the overlap area as a function of the twisted angle and plotted it in Fig. 7(b). The result of the overlap area also shows a minimum at $90^0$ and thus reinforces the fundamental role of the overlap area in all cases of the size of graphite nanofibers.

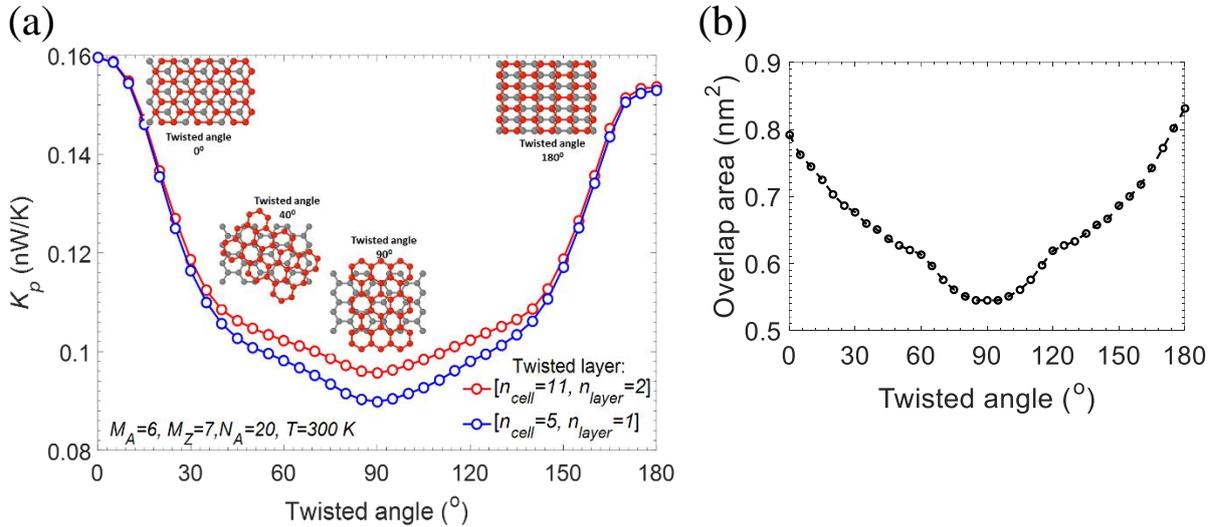

**Figure 7**. (a) Thermal conductance of nanofibers at 300 K with the size $M_A = 6$ ($d_A = 1.21$ nm) $M_Z = 7$ ($d_Z = 0.74$ nm) as a function of the twisted angle. (b) The overlap area of the twisted and fixed layers in a basic cell versus the twisted angle.

## 4. Conclusion

We have studied the phonon transport properties in twisted-layer graphite nanofibers. We demonstrated that in the presence of a twisted layer, the phonon thermal conductance of such systems varies significantly. Interestingly, it showed that the thermal conductance reaches a local minimum at the two angles $\theta_1, \theta_2$ around $90^0$ or at exactly $90^0$ and these crucial angles are independent of the position of the twisted layer and the length of the device, but it depends on the shape of the cross section.

We pointed out that the variation of the thermal conductance stems from the change of phonon modes due to the twisting effect. It was unveiled that the twisting effect impacts strongly the optical modes, in particular, in the frequency range from 70 cm$^{-1}$ to 250 cm$^{-1}$, and it weakly influences the acoustic modes. This phenomenon is similar to that observed with isotope doing in in-plane graphene structures.

We also unveiled that the behavior of the variation of the thermal conductance with the twisted angle is associated directly with the overlap area between the twisted and non-twisted layers.

The magnitude of the crucial angles relies on the size of the nanofibers, and in particular, on the relative ratio between the widths along the zigzag and armchair edges of the nanofibers, thus demonstrating the dominance of the finite-size effect in the variation of the thermal conductance of a twisted graphite nanofiber. Our findings provide fundamental understanding of the thermal transport in twisted-layer graphite nanofibers and also point out potential applications of this effect in areas such as thermoelectrics.

## Acknowledgments

This work was supported by the French National Research Agency via the Placho project grant (ANR-21-CE50-0008). Thanh-Tra Vu also acknowledges the funding by Vietnam National Foundation for Science and Technology Development (NAFOSTED) under grant number 103.01-2018.338